\documentclass[sigconf]{acmart}
\usepackage{url}
\usepackage{hyperref}

\AtBeginDocument{%
  \providecommand\BibTeX{{%
    \normalfont B\kern-0.5em{\scshape i\kern-0.25em b}\kern-0.8em\TeX}}}

\copyrightyear{2021}
\acmYear{2021}
\setcopyright{rightsretained}
\acmConference[BiblioDAP] {}
\acmBooktitle{The 1st Workshop on Bibliographic Data Analysis and Processing}
\acmPrice{}

\begin{document}

\fancyhead{} 
\title{BiblioDAP'21: The 1st Workshop on Bibliographic Data Analysis and Processing}

\author{Zeyd Boukhers}
\email{boukhers@uni-koblenz.de}
\orcid{0000-0001-9778-9164}
\affiliation{%
  \institution{Institute for Web Science and Technologies\\ University of Koblenz-Landau}
  \city{Koblenz}
  \country{Germany}
}

\author{Philipp Mayr}
\email{philipp.mayr@gesis.org}
\orcid{0000-0002-6656-1658}
\affiliation{%
  \institution{GESIS -- Leibniz-Institute for the Social Sciences}
  \city{Cologne}
  \country{Germany}
}

\author{Silvio Peroni}

\email{silvio.peroni@unibo.it}
\orcid{0000-0003-0530-4305}
\affiliation{%
  \institution{Research Centre for\\ Open Scholarly Metadata\\ Department of Classical Philology and Italian Studies \\ University of Bologna, Italy}
  \city{Bologna}
  \country{Italy}
}

\renewcommand{\shortauthors}{Session: Workshop Summaries }


\begin{abstract}
  Automatic processing of bibliographic data becomes very important in digital libraries, data science and machine learning due to its importance in keeping pace with the significant increase of published papers every year from one side and to the inherent challenges from the other side. This processing has several aspects including but not limited to  I) Automatic extraction of references from PDF documents, II) Building an accurate citation graph, III) Author name disambiguation, etc. Bibliographic data is heterogeneous by nature and occurs in both structured (e.g. citation graph) and unstructured (e.g. publications) formats. Therefore, it requires data science and machine learning techniques to be processed and analysed. Here we introduce BiblioDAP'21: The 1st Workshop on Bibliographic Data Analysis and Processing.
\end{abstract}
\keywords{Bibliographic data, Digital libraries, Machine Learning, Data Science}

\ccsdesc[300]{Information systems~Information retrieval}
\ccsdesc[300]{Computing methodologies~Machine learning}
\ccsdesc[500]{Applied computing~Digital libraries and archives}

\maketitle

\section{Introduction}
The aim of the Workshop on Bibliographic Data Analysis and Processing (BiblioDAP\footnote{\url{https://bibliodap.uni-koblenz.de/}}) is to address open challenges in digital libraries and to attract the attention of the research communities to present novel approaches to bibliographic data analysis, processing and understanding. Consequently, we invite the submission of original and high-quality research papers and reports of live demonstrations and prototypes on the following and any related topics: 

\begin{itemize}
    \item Reference Analysis
    \item Citation Network Analysis
    \item Author Name Disambiguation
    \item Scientific data management
    \item Metadata extraction
    \item Plagiarism detection
    \item Bibliographic data quality improvement
    \item Entity linkage in bibliographic data
\end{itemize}

Author Name Disambiguation (AND) is an open challenging problem and its effects are growing as the number of authors sharing the same names arises significantly~\cite{Ferreira2020}. To the best of our knowledge, the last big AND challenge was held in KDD Cup in 2013, which was very successful as several original implemented techniques were presented and available until nowadays. However, since then, the area of Data Science and Machine Learning has significantly evolved together with the availability of computational resources. To fill the gap between the challenges of AND caused by the continuous increase of authors sharing the same name from one side  and the remarkable advancement of artificial intelligence from the other side, we are also organizing a shared task on Author Name Disambiguation~\footnote{\url{https://github.com/BiblioDap/AND-Task}}. To this end, two annotated datasets are released; one for development and one for testing. The workshop invite researchers to submit their results and request those with high accuracy to submit their original research papers, live demonstrations and source codes that tackle this particular problem.

As SIGKDD is a premier Data Science conference that gathers passionate scholars specialized  in data science and its related fields, organizing the workshop in conjunction with KDD 2021 aims to attract the attention of an important part of the data science community that is interested in challenging topics related to bibliographic data.

\section{ Motivation }
The purpose of BiblioDAP'21 is to share knowledge and techniques in data science to be applied to bibliographic data and to overcome its inherent challenges. Data science and machine learning techniques are evolving and progressing significantly being applied to various types of data and for different objectives. BiblioDap' 21 aims to benefit from this progress by bringing together bibliographic data and data science capabilities to overcome the open challenges. The organizers have been managing several projects~\footnote{EXCITE: \url{https://excite.informatik.uni-stuttgart.de/} \cite{hosseini2019}} and infrastructures~\footnote{OpenCitations: \url{http://opencitations.net/}} related to processing and analysing bibliographic data which allow them to have a broad expertise in this discipline and a wide familiarity with the challenges inherent with this data. 

\section{Workshop Setup}

BiblioDAP' 21 consists of one invited talk followed by technical presentations and a panel. The technical presentations are given by the authors of the accepted papers which are reviewed by at least two seasoned reviewers based on the originality and clarity of the paper. 
BiblioDAP' 21 supports open peer review and allows reviewers to disclose their identities and publish their reviews. Using this approach, it is up to the reviewers to decide either to go through an open peer-review process or to keep their names anonymous. The authors of submitted papers were aware that their reviewers may decide to disclose their names and to publish the review openly online in the platform of their choice. In particular, we pointed to specific guidelines (available at \url{https://open-sci.github.io/review/}) that have been devised to foster the adoption of Open Science practices, to publicly acknowledge the effort researchers spend in reviewing, and to enable the reviewers to take public responsibility for the content of the reviews they write to help the authors to improve their work.
Since BiblioDAP' 21 is non-archival, it accepts submissions of papers that are under review to other venues, including KDD' 21. 

\subsection{Keynote}
Our keynote was given by Alberto Laender (Universidade Federal de Minas Gerais, Brazil)\footnote{\url{https://homepages.dcc.ufmg.br/~laender/}} in collaboration with Anderson A. Ferreira (Universidade Federal de Ouro Preto, Brazil) and Marcos André Gonçalves (Universidade Federal de Minas Gerais, Brazil). Alberto and his team talked about ``\textbf{Automatic Disambiguation of Author Names: Foundations, Methods and Open Issues}'' (see their recent book on AND \cite{Ferreira2020}). 

Abstract of their talk: \textit{Author name disambiguation is a well-known hard problem, that has profound impacts on services provided by bibliographic repositories and similar platforms. Despite almost 20 years of research on the topic and efforts such as ORCID, there are still several open issues to be solved. In this talk, we will revisit this problem, presenting an overview and related taxonomy, and elaborate on some methods developed by our own research group that follow distinct approaches and tackle fundamental aspects of the problem such as self training and incremental disambiguation. Finally we will briefly discuss recent approaches and issues still open.}

\subsection{Program Committee}
We appreciate the reviewers’ efforts and would like to thank the members of the program committee for their valuable support.


\section{Related Workshops}
The chairs of BiblioDAP'21 have been involved, in the past, in the organisation of several events and workshops dedicated to scholarly data, including bibliographic and citation data. Among these events there are:

\begin{itemize}
  \item the SAVE-SD workshop series (\url{https://save-sd.github.io/}) is a series of workshops dedicated to technologies aiming at enhancing scholarly dissemination;
  \item the Workshop on Open Citations and Open Scholarly Metadata (\url{https://workshop-oc.github.io/}) has reached its second edition in 2020 and involved researchers, scholarly publishers, founders, policy makers, and opening citations advocates, interested in the creation, reuse, and improvement, of open citation data and open scholarly metadata;
  \item the Bibliometric-enhanced IR (BIR) workshop series (\url{https://sites.google.com/view/bir-ws/home}) tackles issues related to academic search, at the crossroads between Information Retrieval and Bibliometrics \cite{cabanac2020}.
  \item the Scholarly Document Processing workshop (\url{https://sdproc.org/2021/}) covers research and shared task tracks to work on enhancing search, summarization, and analysis of scholarly documents \cite{chandrasekaran2020}.
\end{itemize}


\nocite{*}
\bibliographystyle{ACM-Reference-Format}
\bibliography{acmart.bib}

\end{document}